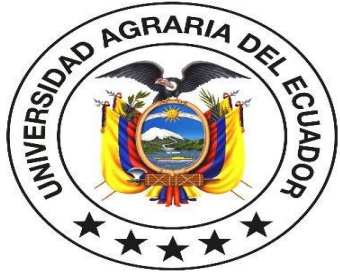

# UNIVERSIDAD AGRARIA DEL ECUADOR

# FACULTAD DE ECONOMÍA AGRICOLA

## CARRERA DE ECONOMÍA

TEMA:

**"INCREMENTO DEL PRECIO DE LOS COMBUSTIBLES Y SU INCIDENCIA EN LOS PRODUCTOS DE LA CANASTA BÁSICA DEL CANTÓN EL TRIUNFO, PROVINCIA DEL GUAYAS".**

AUTORES:

ALVEAR GUZMAN KATHERINE

CAMPOZANO BUELE JENNER

DURAN CAÑARTE PAULETTE

HOLGUÍN CEDEÑO ROGER

MEJÍA CRESPÍN FERNANDO

2022

**RESUMEN**

El objetivo de la presente investigación consistió en analizar el impacto del incremento del precio de los combustibles y su incidencia en los productos de la canasta básica del cantón El Triunfo, en la provincia del Guayas. En el presente estudio se empleó el método de tipo cuantitativo no experimental. La población de estudio se delimitó a las familias del cantón, buscando determinar cómo se vio impactado su nivel de consumo luego del incremento de los combustibles, de ellas se tomó una muestra de 95 personas de manera aleatoria. El instrumento de estudio que se empleó fue encuestas, con enfoque en el poder adquisitivo de las familias con respecto a la canasta básica luego del aumento del precio de los combustibles. Los resultados fueron procesados a través del Alfa de Cron Bach y reflejados en gráficos de pastel, la variable independiente y dependiente que componen nuestro estudio, fueron relacionadas a través de una regresión lineal simple, para determinar si correlacionan entre sí.

**Palabras claves:** Combustibles, Canasta básica, Regresión lineal, Subsidios, Inflación

# ABSTRACT

The objective of this research was to analyze the impact of the increase in the price of fuels and its incidence on the products of the basic basket of the El Triunfo, the province of Guayas. In the present study, the non-experimental quantitative method was used. The study population was limited to the families of the town, seeking to determine how their level of consumption was impacted after the increase in fuels. Just 95 people were randomly taken. The study instrument that was used was surveys, with a focus on the purchasing power of families with respect to the basic basket after the increase in fuel prices. The results were processed through Cronbach's Alpha and reflected in pie charts. The independent and dependent variable that make up our study, were related through a simple linear regression, to determine if they correlate with each other.

# ÍNDICE DE CONTENIDOS

# INTRODUCCIÓN

## Caracterización del Tema

En el presente artículo se conocerá cuál ha sido el impacto de la subida de los combustibles, en los precios de los productos de la canasta básica. En los últimos años el precio del barril del petróleo ha sufrido cambios debido a la volatilidad de los mercados internacionales y de factores atípicos como la pandemia causada por el virus (SRAS-CoV-2) y la invasión de Rusia a Ucrania. A la presente fecha el precio del barril de crudo West Texas Intermédiate ronda los $107, 73 USD, mientras que el de Brent se encuentra en $110,27 USD.

En Europa se ha registrado un alza del precio de los combustibles, en países como España en el precio medio del litro de gasolina de 95 octanos ha pasado de 1,59 a 1,68 euros, lo que supone un alza del 5,5%, en Alemania el precio del litro de la gasolina y el gasóleo ya supera los dos euros, lo que representa un incremento del coste de la gasolina de un 14,5% y Holanda presenta un aumento del 8,8% para la gasolina de 95 octanos y del 12,9% para el gasóleo. Esto se ha dado por la guerra entre Rusia y Ucrania y los problemas de suministro que están afectando a toda Europa.

En Ecuador se ha dado un aumento progresivo de los precios de combustibles desde la eliminación de los subsidios, debido a que nuestro país importa el 66% de los hidrocarburos que consumimos y a que el precio del combustible está muy ligado al costo del petróleo crudo, que es la materia prima de la cual se refina.

Todos estos factores se vieron reflejados en afectaciones para la sociedad debido a su impacto directo a los productos de la canasta básica, en España los productos derivados de los aceites vegetales, cereales y la carne han sido los principales responsables de esta subida histórica. Para el mes de marzo de 2022 aumentaron en un 17,1% respecto a febrero de ese mismo año, este mismo problema afecta a toda Europa, en el mes de febrero Francia cerró con una inflación del 3,6% frente al mismo mes del año pasado.

Para el año 2022, en Ecuador el precio de la canasta familiar básica (CFB) se ubica en $725,16 USD, un aumento de $13,15 USD con respecto al año 2018 donde el precio se situaba en $712,03 y un incremento anual de 1,84% respecto a febrero de 2022.

Los productos que más se han encarecido en el país son el aceite vegetal con un aumento del 35% en el mes de marzo de 2022 con respecto a marzo del 2021, también la manteca, la margarina y la harina, que han incidido directamente en el aumento de la canasta básica familiar y ha agravado la brecha social y crisis económica que se vive en el país.

**Planteamiento de la Situación Problemática**

En el presente artículo estableceremos la relación que existe entre el incremento de la subida de los precios en la canasta básica por consecuencia de la eliminación de los subsidios en los combustibles en nuestro país y otros factores externos que afectan el mercado internacional de los combustibles.

Dentro de los factores que han hecho incrementar el valor de los combustibles en el mercado internacional, debemos mencionar la crisis que se vive en Europa por la Guerra entre Rusia y Ucrania, además de mencionar que en nuestro país desde el 2019 se eliminaron los subsidios a los combustibles lo que nos hace pagar el precio internacional de los mismos.

Al hablar de los combustibles debemos hablar del diésel el cual se utiliza principalmente para el transporte de carga y de pasajeros, por lo que al incrementarse su costo genera un efecto inflacionario directo en los productos de la canasta básica, recordar que el valor por galón se incrementó en un 100% tras aplicar eliminación de los subsidios.

Analizaremos los efectos que han sufrido los consumidores tras el incremento de algunos productos de la canasta básica, entre ellos, el aceite vegetal, harina y algunos vegetales, son de los que más han sufrido un alza en sus precios, tener presente que nuestro país atraviesa una crisis de desempleo, en donde la tasa de desempleo es del 6.2% lo que afecta a las familias al momento de adquirir los productos de primera necesidad.

**Justificación e importancia del estudio**

Ecuador ha resaltado entre los demás países de la región por los importantes subsidios que se le otorgan a los combustibles estableciendo una relación directa entre la comercialización y el precio de compra, la primera vez que se implementó esta medida de subsidiar a los hidrocarburo fue en el año 1974, se empleó como una política social y compensatoria, declarando que esto iba a incentivar el consumo principalmente en las familias de ingreso medio y bajo, la economía ecuatoriana depende de los ingresos que tenga de la exportación de petróleo si bien es cierto representa un gran ingreso en el presupuesto general del Estado, pero la subida de precio por algunos factores tales como la guerra, también le afecta en gran medida ya que importa el 60% de los combustibles para satisfacer la demanda local, y subsidiarlos provoca desequilibrio en el déficit fiscal, por lo cual deja que el ciudadano cubra ese porcentaje que cubría el estado anteriormente, esto representó un alza en su precio y como consecuencia la subida de coste transporte para la comercialización de algunos productos.

El ingreso familiar tiene una relación directa con la canasta básica porque dependiendo de sus ingresos puede gozar de una canasta básica adecuada, el problema principal es que el precio de la canasta básica familiar supera al sueldo promedio de los habitantes la cual pierde poder adquisitivo.

**Delimitación del problema**

En este trabajo de investigación se analizará el incremento del precio de combustibles y su incidencia en los productos de la canasta básica del cantón el Triunfo, provincia del Guayas, en el periodo 2018-2022.

**Formulación del problema**

¿Cómo influye el incremento del precio de combustibles en los productos de la canasta básica del cantón el Triunfo, provincia del Guayas, en el periodo 2018-2022?

**Objetivos**

**Objetivo general**

Analizar el impacto de la subida de combustibles en los productos de la canasta básica en el Cantón El Triunfo, Provincia del Guayas.

**Objetivos específicos**

- Describir la canasta básica y los efectos sociales que ocasiona el aumento de los precios.
- Establecer la relación entre el precio del combustible Diésel y el precio de la canasta básica, en el periodo seleccionado, mediante una regresión lineal simple.
- Determinar la afectación directa que tuvo el alza de combustibles en el consumo básico de las familias del Cantón El Triunfo-Guayas.

**Hipótesis**

El incremento del precio de los combustibles genera un efecto inflacionario directo en los productos de la canasta básica, aumentando el costo de vida.

# CAPÍTULO I
## Marco Referencial
### 1.1 Estado del Arte

El mercado petrolero se encuentra alejado de ser un mercado de competencia perfecta, esto según Picón 2016, como se citó en (Aguilar, 2019), en su trabajo "el papel del crudo en la economía, factores que influyen en el precio", se presenta debido a las ineficiencias desde el lado de la oferta. Esto ayuda a comprender por qué es complejo desarrollar un modelo eficiente para explicar el equilibrio de este mercado.

Según (Ramírez & Dionicia, 2020) ,en su tesis "Subsidios a los combustibles y la canasta básica en América Latina, período 2015- 2019", Ecuador y Argentina no tienen la capacidad de satisfacer sus necesidades nacionales de gasolina, por lo que importan gasolina y diésel para abastecer a la demanda local. En el caso de Ecuador, a pesar de la existencia de leyes de hidrocarburos, este subsidio implícito no se registraba en las cuentas nacionales (Mendoza, 2014). Aunque representó alrededor del 8,38% y el 3,01% del presupuesto general del Estado en los últimos veinte años. Durante el primer auge petrolero de Ecuador, los gobiernos dirigidos por militares aumentaron los subsidios a los combustibles en un esfuerzo por beneficiar a los grupos sociales marginados.

Los subsidios a los combustibles reflejan una carga estatal la cual nuestro país no puede mantener, pero su afección es directa a la economía familiar como explica (Chicaiza, 2019), en su investigación "Las políticas de eliminación en los subsidios de los combustibles fósiles y su relación con la inflación del Ecuador", donde comenta que el tema de la eliminación de los subsidios a los combustibles siempre ha sido un tema de disputa y un factor para el crecimiento de la inflación ya que causa incidencia en los hogares ecuatorianos ya que son los consumidores principales de los productos derivados del mismo.

Bowen y Chimbolema (2021), en su trabajo de titulación "Incremento del precio del combustible diésel y su incidencia en el precio de la canasta básica en Guayaquil período 2015-2020", en dicho trabajo de titulación se hace relación al incremento de los precios de la canasta básica y el incremento que se da en los costes del transporte de carga y pasajeros, lo que termina desencadenando un efecto inflacionario en la canasta básica de nuestro país.

Por otra parte (Bure, Guerrero, Aguirre, & Gaona, 2021) en su investigación de la "evolución del costo de la canasta básica en los diferentes periodos presidenciales 2000-2019", se llegó a la conclusión de que en la última década, las familias mejoraron su calidad de vida y un incremento de los ingresos logrando acceder a más de 50% de los productos de la canasta básica disminuyendo la brecha y lograr una mayor equidad, mediante políticas implementadas para mejorar el ahorro, vivienda y salud, aunque es necesario destacar que gran parte de la población económicamente activa no tiene ingresos estables y por lo tanto no puede acceder a este conjunto de productos.

### 1.2 Bases Científicas y Teóricas de la Temática

#### 1.2.1 Economía de Bienestar.

La economía de bienestar es una rama de economía que estudia el método para orientar el sistema económico hacia el bienestar social y así poder elegir aquel sistema que más promueva el desarrollo humano y social.

Las personas somos conscientes de nuestras necesidades como explica (Riascos et al., 2020) El bienestar en la economía incorpora elementos de diferentes escuelas que la estudian, Arrow (1951) mediante su teorema de imposibilidad fundamenta el análisis de las elecciones de los individuos sobre sus valores y no en la simpleza de sus gustos, por lo que define que el bienestar no se logra por medio de la dinámica del mercado.

Esto establece que nosotros hacemos elecciones basandonos en nuestros deseos pero analizando antes nuestros valores, y de esta manera la economía de bienestar busca estos valores y deseos sociales ante el desarrollo en comunidad.

##### 1.2.1.1. Objetivos de la Economía de Bienestar

Entre las metas que sigue la economía de bienestar está el descubrimiento de un sistema económico que intente maximizar los recursos, con el propósito de aumentar la paz social.

Por medio del análisis de los sistemas económicos, el propósito de esta rama del pensamiento económico se reúne en un objeto de análisis como es la maximización de la producción con unos los recursos limitados dados, optimizando el reparto de los bienes y servicios elaborados, su objetivo primordial es el aumento del confort social.

Entonces podemos definir al bienestar social como el bienestar del conjunto total de la sociedad. Existen, por tanto, dos formas de medir la suma del bienestar de una población. Que son el método ordinal, desarrollado por el economista Wilfredo Pareto. Y también el método cardinal, el cual se basa en la medición del valor en términos monetarios.

### 1.2.2  ¿Qué es Inflación?

La inflación es un crecimiento generalizado en los costos de los bienes y servicios de una economía a lo largo de un tiempo determinado.

Como explica (Atucha et al., 2018) La inflación es un fenómeno que cuenta con tres características, primero el aumento de precios a precios no altos, segundo aumento de precios recurrente y el último establece que prácticamente todos los precios tienen que incrementarse.

#### 1.2.2.1.  ¿Cómo Combatir la Inflación?

En ciertos territorios los bancos centrales fijan tasas de interés y controlan la proporción de dinero que circula en la economía. Una forma clásica es subiendo la tasa de interés, lo que disminuye la proporción de dinero circulando, pero aquello puede fomentar el desempleo y estancar el incremento de la economía. Otros sostienen la teoría de combatirla fijando la tasa de cambio de la moneda local, ante las divisas fuertes, en especial el dólar. Ciertos plantean que lo más adecuado es minimizar los impuestos y dejar que la tasa de cambio flote según las fuerzas del mercado

#### 1.2.2.2    Ventajas de la Inflación

De forma directa pues los bienes poseen costos más bajos, y de una forma indirecta debido a que inflaciones negativas involucran comúnmente tasas de interés además negativas, por lo que clientes y organizaciones tienen la posibilidad de lograr superiores condiciones para su financiación.

### *1.2.3. ¿Qué son los Subsidios?*

Un subsidio o incentivo del gobierno es una manera de ayuda o apoyo financiero que se prolonga a un sector económico para fomentar determinadas políticas económicas y sociales. Puede tratarse de una prestación económica de una duración definida en la época o no.

Como expresa (Escribano, 2019) en su escrito "Ecuador y los subsidios a los combustibles" El subsidio a los combustibles ha sido introducido en Ecuador en 1974 por el sistema militar en un entorno de costos al levanta del petróleo y de crecimiento de la producción doméstica que proporcionaba ingresos crecientes. La bonanza petrolera se usó para reforzar al sistema sin que los subsidios fueran reducidos significativamente una vez que el caso económico ha cambiado, pues los gobiernos posteriores fueron conscientes de la impopularidad de retirarlos.

En la segunda mitad de los años 90, los sucesivos intentos de eliminarlos encontraron una intensa contraposición famosa y provocaron serios episodios de inestabilidad política, a pesar de lo que fueron casi erradicados y sustituidos por transferencias directas a los domicilios más pobres, que luego se convirtieron en el en la actualidad vigente Bono de Desarrollo Humano (BDH).

No obstante, a lo largo de una serie de crisis económicas y bancarias, las reducciones de los subsidios energéticos contribuyeron a la caída de diversos mandatarios.

El mandatario Correa mantuvo el grado de los subsidios, sin embargo, confrontado con una situación económica cada vez más complejo, su sustituto Moreno admitió (entre otras reformas ya citadas) una reducción gradual de los subsidios de ciertos combustibles para obtener la ayuda financiera del Fondo Monetario Internacional. En 2018 se aplicó una reducción de los subsidios a la gasolina súper y en 2019 fueron al final destruidos, mientras tanto que se redujeron los subsidios a la gasolina habitual.

### *1.2.4. ¿Qué es la Canasta Básica?*

El término canasta básica se lo puede definir como el conjunto de productos y servicios considerados esenciales para la subsistencia y bienestar de una familia durante un lapso de tiempo establecido, esta incluye alimentos, vestimenta, higiene, salud, transporte, entre otro (Guzmán, 2020).

### 1.2.5 El Salario Básico en Ecuador.

El salario básico o también conocido como salario mínimo, se define como el pago mínimo mensual que debe hacer el empleador a sus trabajadores por un trabajo que hayan efectuado durante en un periodo de tiempo ya establecido (Organización Internacional del Trabajo).

Actualmente el SBU en nuestro país se encuentra en 425 dólares americanos en este año 2022.

### 1.2.6 Eliminación de los Subsidios y la Inflación en el País.

La eliminación de los subsidios se da parcialmente en octubre del año 2019 por medio de un decreto presidencial, medida que luego de unos días fue suspendida por las protestas en el país, sin embargo, luego de unos meses volvió a ser aplicada hasta la actualidad que se mantiene con los precios fijos tanto en diésel y gasolina súper y se respeta el precio internacional en la gasolina súper, llegando a precios de 5 dólares por galón. (España, Ecuador elimina los subsidios a la gasolina para corregir sus estrecheces fiscales, 2019)

Anteriormente ya mencionamos que era inflación, ahora vamos a hablar acerca de la inflación en nuestro país, actualmente en el mes de abril del presente año se vio reflejada porcentualmente en 2,89%, esto quiere decir que se da un incremento del 0,59% con respecto al mes anterior. (Coba, 2022)

### 1.2.7 Recuento de los Subsidios a los Combustibles en el País.

Los subsidios a los combustibles en el Ecuador, eliminados en el año 2019, tienen su origen en la década de los 70, tal como señala (Espinoza & Viteri, 2019) en su "Análisis económico a la eliminación de los subsidios a la gasolina súper en el Ecuador", se crearon con el propósito de estimular a los sectores empobrecidos en áreas como salud y alimentación.

Los subsidios han ganado tanta relevancia a nivel nacional que en el año 2000 formaron parte de la política monetaria del gobierno, también se han tornado razón de presión social. En el año 2003 se establecen los subsidios a los combustibles que no habían sido modificados hasta el año 2019, representando para el Estado una carga de millones de dólares en el gasto público.

De acuerdo al (Ministerio de economía y finanzas, 2022) en los últimos 12 años se destinaron 45 mil millones de dólares para financiar diversos subsidios, siendo el de los combustibles el de mayor peso para el estado representando una cifra anual de 1.707,04 (millones) USD.

Durante 2011-2018, los combustibles que más subsidió el Estado fueron el Diésel e importación de nafta para atender la demanda de producción de gasolina de alto octanaje como súper y extra. 2013 fue el año de mayor inversión en subvenciones, mientras que desde 2016, se observó una disminución significativa. En el año 2015 se suspendieron temporalmente los subsidios al Diésel y durante los próximos años se continuó con una búsqueda por la disminución de estos, en el gobierno de Rafael Correa se propuso crear hidroeléctricas para así no depender de la energía térmica. Esto redujo la importación de Diésel. (Espinoza & Viteri, 2019)

En el año 2019 se propuso la eliminación de los subsidios a los combustibles, lo que provocó una revuelta social de varios días que finalizó en un acuerdo mutuo entre el gobierno y los colectivos sociales.

Finalmente, en el año 2020 se eliminan los subsidios a los combustibles que fueron reemplazados por un sistema de precios conocido como bandas fluctuantes. Así, el precio piso para la gasolina Extra será de USD 1,75 y podrá fluctuar hacia arriba máximo en 5%, al igual que la banda del precio base del Diésel que es USD 1 podrá fluctuar en el mismo porcentaje.

### 1.2.8 Composición de la Canasta Básica en el Ecuador.

En el país, la canasta básica familiar se compone de 75 productos y servicios. Según (Suarez, 2022), en su artículo denominado "La canasta familiar está más que cubierta, pero para pocos", el 2022 comenzó con precios altos de manera generalizada. Enero de este año empezó con una inflación mensual positiva de 0,72%, la más alta desde 2014. La sección de consumo que más contribuyó al incremento de la inflación mensual fue la de Alimentos y bebidas no alcohólicas.

En palabras de (Freire, Mayorga, Vayas, & Sánchez, 2020), en su trabajo de investigación denominado "La canasta básica ecuatoriana, panorama general", según datos del (INEC, 2020), en enero del año 2013 el 92% de los ecuatorianos pudieron costear la canasta familiar básica (CFB) que fue de $589,39, en tanto que, en el año 2016, solo el 67% de las familias pudo hacerlo.

Para llevar a cabo el análisis de la CFB el INEC considera los bienes y servicios indispensables para satisfacer las necesidades de una familia de 4 miembros (madre, padre y dos hijos), en dónde 1,6 percibe ingresos y ganan la remuneración básica.

La CFB está conformada por cuatro clases de productos, alimentos y bebidas, viviendas; indumentaria y misceláneos.

### 1.2.9 Evolución del Precio de la Canasta Básica en Ecuador.

La CFB puede cambiar de acuerdo con las costumbres, hábitos, recursos de las familias y región, debido a esto es una canasta básica promedio que puede variar en los elementos que la componen presentados en el apartado anterior.

En el año 2000, bajo la presidencia de Jamil Mahuad Ecuador se convertía en el primer país latinoamericano en adoptar el dólar como moneda nacional. Esto trajo consigo consecuencias políticas y sociales que desencadenaron principalmente en una crisis económica que provocó un aumento del desempleo, que afectó principalmente a la clase más baja, tal como señalan (Brito , Quito, Rodríguez , & Uriguen, 2021), en su artículo denominado "Evolución del precio de la canasta básica del Ecuador. Análisis del periodo 2000 –2019". Esto trajo consigo un aumento de la pobreza, el porcentaje de personas que no podían acceder a la canasta básica pasó de 46% en el año 1998 a 66% en el año 2000, principalmente debido a la inflación los productos de gran consumo llegaron a aumentar hasta en un 70% de su precio.

A partir de todos estos cambios, en enero de 2021 el ingreso promedio de las familias se estimaba en 80,30USD, para ese mismo mes la CFB costaba alrededor de 170USD.Para el año 2003 la economía ecuatoriana había decrecido en un -0,3%, luego ingresó a un periodo de estabilidad, pasando el nivel de inflación de 12,5% en 2002 a 2,7% en 2004. A pesar de esto el ingreso de las familias no alcanzaba para adquirir la CFB en su totalidad la cual, para abril de 2005, tenía un valor de 425,12USD.

Durante el periodo 2006-2016, según él (Ministerio de inclusión económica y social, 2022) la pobreza en el país disminuyó en un 16,5%, esto se vio reflejado en un incremento en los ingresos familiares siendo así que para el 2015 alcanzaban a cubrir casi la totalidad el costo de la CFB que era de 681USD.

En la actualidad el costo de la canasta básica es de 725,27USD y la inflación en marzo del presente año llegó a 2,64% según él (INEC, 2020).

### 1.2.10 Inflación a los Productos de la Canasta Básica.

Desde hace algo más de una década, el tema de la inflación ha dejado de ser un tema de analistas, para convertirse en un hecho de la vida cotidiana, un componente esencial de la conciencia de cada ciudadano, una causa de angustia y preocupación.

En efecto, luego de la crisis económica de los años ochenta, todos los latinoamericanos, sin excepción, comenzamos a descubrir que nuestros salarios no alcanzaban para comprar lo que necesitábamos; Día tras día, los precios de las materias primas aumentaron de manera constante e irresistible; que las empresas y el propio estado han adoptado políticas de recortes presupuestarios, descuidando los programas de protección social, e incluso poniendo a muchos trabajadores en un desempleo impredecible; que incremento el costo de los servicios básicos como agua, electricidad, teléfono y combustible; que ya no éramos capaces de ahorrar y que el crédito había llegado a niveles peligrosos; El resultado de toda esta situación fue que las clases medias se empobrecieran y los pobres pasaran a una condición crítica (Schuldt & Acosta, 2013).

Históricamente, la inflación comenzó a subir con fuerza durante la temporada cacaotera, debido a una serie de factores, como la caída del precio internacional del cacao conocido como el grano de oro, que se ha atribuido a una serie de hechos como la Primera Guerra Mundial. La plaga y la supresión de nuevos competidores en el comercio del cacao.

Porque la economía del país depende principalmente de la producción de cacao, que supone dos tercios de las exportaciones del país, así como de una fuerte caída de los precios del azúcar para poder competir en el mercado internacional y reducir las importaciones. Para crear un indicador de inflación en ese momento, los artículos de primera necesidad tomados por la sección comercial de los diarios de Guayaquil, El Universo, El Comercio, El mercurio, El ecuatoriano, El Telégrafo, El Grito del Pueblo Ecuatoriano.

Los productos que se utilizaron para medir la inflación en esos momentos fueron:

- 5 libras de arroz
- 5 libras de azúcar
- 1 libra de café molido
- libras de frejol
- 1 libra de manteca
- 2 libras de papa
- huevos
- libras de carnes sin hueso

Los estudios indican que en los primeros 14 años que van desde el 1910 a 1923 alcanzó el 66,81%, alcanzando su punto más alto en el año 1922 con el 75,67%. (Salazar & Zurita, 2017)

La canasta básica que nació a principios del siglo XX es la idea del químico británico Seebohm Rowntree, quien a través del proceso de estudiar la cantidad exacta de proteínas y calorías para que el cuerpo humano funcione correctamente con la finalidad de estructurar el problema de la pobreza de los obreros en la ciudad de Nueva York, Estados Unidos.

Desde el nacimiento de la República del Ecuador en 1830, en la primera Asamblea Constituyente, surge la necesidad de contar con información estadística sobre el desempeño de los representantes de las tres divisiones más importantes como lo son Azuay, Guayas y Quito; Esto se haría por censo. Aunque en 1830 y 1973 se establecieron varias organizaciones relacionadas con el censo; El Instituto Nacional de Estadística y Censos (INEC) no fue creado hasta 1976, mediante el Decreto 323, al fusionarse el Instituto Nacional de Estadística, la Oficina Nacional de Estadística y el Centro de Análisis. (INEC, 2017)

El INEC, es la institución encargada de realizar los estudios de la canasta básica familiar, todos los meses analiza el comportamiento de los demandantes de los productos que componen la canasta, al igual que realiza el censo poblacional y estudia el comportamiento inflacionario del país.

Supermercados, mercados, pulperías, tiendas y otros lugares en El Triunfo venden productos de la canasta básica, los cuales han aumentado significativamente a lo largo de los años. La inflación es causada por un aumento continuo de los precios, que generalmente se mide por el índice de precios al consumidor. La tasa de inflación es el cambio porcentual en el precio de los bienes y servicios durante un período determinado.

La presencia de inflación significa un aumento continuo de los precios de los bienes en general, lo que perjudica el poder adquisitivo de la población, reduce su poder adquisitivo y afecta su calidad de vida. El aumento de precios provocado por la inflación se puede medir mediante diversos indicadores que reflejan el crecimiento porcentual medio de la canasta familiar ponderada en función de lo que se pretenda medir.

El índice más utilizado para medir la inflación es el Índice de Precios al Consumidor, conocido como IPC, que indica el cambio porcentual en el precio promedio de los bienes y servicios que recibe un consumidor; para dos periodos de tiempo, tomando como referencia la canasta familiar. El IPC es el indicador más utilizado, aunque no se considera una medida absoluta de la inflación, ya que solo representa la variación real de los precios de los hogares y familias. (Rivera & Villacis, 2019)

### 1.2.11 Incidencia del Conflicto Bélico a Productos en Ecuador.

Para empezar, entendemos por conflicto bélico o armado "cualquier enfrentamiento que libran grupos de diversa índole (como fuerzas militares regulares o irregulares, milicias, grupos armados de oposición, paramilitares, étnicos o religiosos, utilizando armas u otros medios de destrucción, provocando más de 100 bajas en un año". (Icaria, 2005)

El conflicto de Ucrania está causando daños colaterales en muchas otras partes del mundo. Los precios del trigo y otros cultivos básicos se están disparando, amenazando a los países en desarrollo más pobres.

La economista Wilma Salgado advirtió que, con el conflicto entre Rusia y Ucrania, Ecuador tendrá un fuerte impacto en las importaciones, ya que estos países son grandes productores de granos. "Producen el 29% de las exportaciones mundiales de trigo, el 20% de maíz, el 80% aceite de girasol".

Señaló que en Ecuador no se produce trigo, sino que se importa el 97% de lo que se consume nacionalmente, por lo que si se suspenden o restringen las importaciones desde Rusia la gente verá subir el precio del trigo. "Recientemente vimos a un grupo de panificadores pidiendo un aumento sustancial del precio del pan popular del 0.15 a 0.25 centavos, pese a que aún no se siente el impacto del conflicto".

A criterio de la especialista también subirá el precio del maíz, lo cual quiere decir que se encarecerán los costos del balanceado para la alimentación de los animales como los avícolas: "Entonces vemos que hay múltiples efectos negativos sobre la economía ecuatoriana".

"El rubro se está devaluando entonces provoca que los productos ecuatorianos se encarezcan frente a otros exportadores que pueden devaluar su moneda". (Davila, 2022)

Según la opinión del gobierno expresada por el ministro de Industria, Julio José Prado, la guerra podría poner en riesgo unos $1.000 millones este año, por el potencial impacto en las exportaciones del país a Rusia. Para productos como plátanos, camarones, flores, pesca y cacao.

Las complicaciones en el corto plazo no se quedan ahí. Según Rodrigo Gómez de la Torre, analista agropecuario, también podríamos enfrentar la escasez de productos imprescindibles para el sector agrícola, como la urea, puesto que Rusia es el principal exportador de este químico utilizado por todos los agricultores y ganaderos a nivel nacional para fortalecer la productividad del suelo. Además, existe un súbito incremento en el valor de varios productos primarios como el trigo. El resultado de estas variaciones puede terminar siendo el incremento de los costos de producción, encareciendo los precios de varios productos de la canasta básica.

El shock externo producto del conflicto bélico puede golpear principalmente a los estratos más populares del país. Según el economista Byron Villacís, los incrementos de los precios afectan mayoritariamente a las familias de estratos medios y populares, quienes en proporción pueden terminar gastando hasta un 43% de sus ingresos en categorías que sufrirán subida de costos como la alimentación y el transporte(Gronneberg, 2022).

### 1.2.12 Afección de la Eliminación de los Subsidios en el Desempleo.

Un estudio realizado por el Banco Interamericano de Desarrollo (BID) identifica que la eliminación de los subsidios a los combustibles, sin compensación o focalización, afecta de manera negativa a los hogares más vulnerables del país, por lo que plantea algunas alternativas: aumentar el valor y/o a los beneficiarios del Bono de Desarrollo Humano, vales de comida y transporte o, incluso, que los sectores más pobres del país accedan al transporte público gratuitamente.

Aparte de los problemas fiscales y sociales, el Ecuador también se enfrenta a problemas relacionados con el cambio climático. Si bien no es una tarea fácil, un estudio del BID sugiere que reemplazar cuidadosamente los subsidios de energía con un mejor gasto en protección social podría ser una forma en que los gobiernos ecuatorianos progresen en los tres frentes: fiscal, social y ambiental.

El análisis técnico sobre la eliminación de los subsidios indica claramente su ineficiencia, regresividad y alto costo. Lo que el gobierno ahorraría al eliminar los subsidios a los combustibles fósiles podría usarse para reducir los déficits y la deuda soberana, o financiar inversiones en educación, salud o infraestructura. Además, quitar los subsidios también ayudaría a reducir el consumo de combustibles fósiles que contribuyen a la contaminación del aire y representan una gran fuente de emisiones de gases de efecto invernadero.

El problema es que, al reducir los subsidios a los combustibles fósiles, aparte ser un desafío político, los hogares pobres y vulnerables en Ecuador se verían afectados negativamente, como lo indican los resultados del estudio del BID. Eliminar los subsidios a la electricidad y al diésel les costaría alrededor del 2% de sus ingresos. Incluso si no consume diésel directamente, los ecuatorianos pobres son vulnerables al aumento de los precios de los alimentos y el transporte público debido a la eliminación este subsidio.

Algunas alternativas a la eliminación de los subsidios a la gasolina y la electricidad:

- Aumentar el monto del Bono de Desarrollo Humano.
- Expandir el Bono de Desarrollo Humano a nuevos beneficiarios.
- Transporte público gratuito para los ecuatorianos más pobres.
- Emisión de vales de comida a grupos vulnerables.

Estas propuestas dejarían a los ecuatorianos pobres en mejor situación económica, en comparación con la situación actual, y le ahorrarían al gobierno alrededor de $ 1,6 mil millones por año. Además, este paquete podría ganar impulso político, ya que el factor político incide fuertemente en la eliminación de estos subsidios(Suárez, 2021).

### 1.2.13 Cantidad de Combustible Importado en el País.

Para empezar, definamos que es combustible, según (Westreicher, 2020) es cualquier materia capaz de liberar energía cuando se oxida de forma violenta con desprendimiento de calor poco a poco, este insumo se utiliza para diversos procesos, como el funcionamiento de maquinaria, la generación de electricidad o el embalaje de medios de transporte.

El ecuador gasto 2.346 millones de dólares en importar combustible entre enero y abril de 2022 lo que representa un 68% más que el año anterior en los mismos meses, aunque el ecuador es un país petrolero no produce lo suficiente para satisfacer la demanda local debido a que solo produce 218.000 barriles por mes representando el 80% del consumo local por lo que importa 267.000 barriles mensuales para cubrir la demanda restante (Orozco, 2022).

### 1.2.14 Encarecimiento del Costo de Vida en Ecuador.

De acuerdo con (traders.studio, 2022) El costo de vida es la cantidad de dinero requerida para cubrir los gastos básicos como vivienda, alimentación, vestimenta, impuestos y atención médica durante un determinado período de tiempo. Este indicador se usa a menudo para comparar el costo de vida en una ciudad con otra. Gastos de manutención relacionados con el salario. Si los gastos son altos en una ciudad, entonces los salarios deben ser más altos para que la gente pueda vivir en esa ciudad.

El costo de vida se ha encarecido también para los ecuatorianos debido a la inflación que golpea el bolsillo del ciudadano a través de la educación (0,117 % de IPC anual), la vivienda (0,083 %) y la salud (0,069) la cual se incrementó este año. Esta inflación se dan por factores como la subida de precios internacionales del petróleo aparejada a la retirada de los subsidios estatales y la reactivación de la vida una vez que la vacuna contra la covid-19 llegó a millones de ecuatorianos, y debido a eso el salario mínimo que es 425 no alcanza para cubrir dichas necesidades (España, El costo de la vida se encarece en Ecuador ante la subida del petróleo y la reactivación económica, 2021).

### 1.2.15 Impacto Social ante Eliminación del Subsidio a los Combustibles en el País.

El impacto social corresponde a los cambios que experimenta una persona, grupo o comunidad como resultado del desarrollo de una determinada actividad, proyecto, programa o política y afecta la condición de las personas a corto o largo plazo, estos cambios pueden producirse directa o indirectamente y pueden ocasionar efectos positivos o negativo (esimpact, 2021).

La eliminación de los subsidios a los combustibles afectaría el nivel de consumo porque existe una relación directa entre precio de venta de combustible y economía doméstica lo cual los obligaría a priorizar sus gastos intrafamiliares redistribución de los recursos económicos para suplir sus necesidades que notoriamente se verán incrementada por el impacto social que representaría esta decisión política. Esto da como resultado el encarecimiento del costo de vida y limitando más aun su ya vulnerable poder adquisitivo ante una realidad económica cada vez más limitada en la satisfacción de sus necesidades (Ruiz, 2021).

### 1.2.16 Regresión Lineal

Los modelos de regresión describen la relación entre variables ajustando una línea a los datos observados. Los modelos de regresión lineal usan una línea recta, permite estimar cómo cambia una variable dependiente a medida que cambian las variables independientes. (Amat, 2016)

### *1.2.17 Correlación Lineal*

Esta herramienta cuantifica de qué manera están relacionadas dos variables, es decir, es el número que mide la intensidad de la relación entre dos variables. (Amat, 2016)

### 1.3 Fundamentación Legal

Las bases legales de la investigación están representadas en la Constitución de la República del Ecuador (2008).

### *1.3.1 La Constitución de la República:*

La constitución de la República del Ecuador en su Artículo 313 establece que, el Estado es el que administra, regula, controla y gestiona los sectores estratégicos, de conformidad con los principios de sostenibilidad ambiental, precaución, prevención y eficiencia. Adicional en este mismo artículo se detalla la lista de los sectores que son considerados estratégicos la energía, los recursos naturales no renovables, el transporte y la refinación de hidrocarburos (CONSTITUCIÓN DE LA REPÚBLICA DEL ECUADOR, 2008).

Por otro lado, el artículo 317, dictamina que los recursos no renovables no pertenecen definitivamente al estado, por lo que su gestión se hace responsable en la conservación de la naturaleza, cobro de regalías, otras contribuciones y ayuda a minimizar los impactos negativos ambientales, culturales, social y económico.

En cumplimiento al Decreto Ejecutivo No.1054 de 19 de mayo de 2020, mediante el cual se reforma el "Reglamento de Regulación de Precios de Derivados de Petróleo", y en consideración al nuevo "sistema de precios de mercado para la comercialización de combustibles", a partir del 11 de julio, la Empresa Pública Petroecuador aplicará la fijación y publicación de precios en terminal de las gasolinas Extra, Eco país (extra con etanol) y Diésel para los segmentos automotriz, camaronero, atunero y pesquero, a través del mecanismo técnico de banda móvil del más/menos el 5% (PETROECUADOR, 2020).

Decreto 883: eliminación de los subsidios de combustible. Decreto883: derogado, sustituido por art.894.

## CAPÍTULO II

## Aspectos Metodológicos

### 2.1 Métodos

**Método Deductivo**: En la presente investigación se utilizó el método deductivo ya que se ha usado la lógica para llegar a una conclusión mediante un conjunto de afirmaciones que se dan por verídicas. Durante la elaboración de este artículo científico se ha ido de una premisa general, para llegar a una conclusión particular que nos permita probar o negar la hipótesis.

### *2.2 Modalidad y Tipo de Investigación*

Para el desarrollo de esta investigación se recabó información de distintas fuentes de consulta, además de investigaciones ya existentes de diferentes autores que fueron citadas y referenciadas debidamente.

#### *2.2.1 Modalidad no Experimental.*

La elaboración del presente artículo se realizó mediante la modalidad no experimental, debido a que se han empleado conceptos, sucesos, contextos, variables en las que no se ha tenido intervención directa, si no que se han observado situaciones ya existentes.

#### *2.2.2 Investigación Descriptiva.*

La investigación descriptiva es de tipo observacional, debido a que no se tiene ninguna influencia sobre las variables presentadas en el estudio. Hace referencia a las preguntas de investigación, análisis de datos y diseño del estudio.

#### *2.2.3 Método Cuantitativo.*

El método mediante el cual se recolectará la información necesaria para dar respuesta a los objetivos del presente trabajo será el cuantitativo, debido a que se implementará una encuesta que nos permita medir y presentar resultados en datos numéricos.

### 2.3 Variables

Las variables de investigación son diferentes propiedades o características de los objetos, cosas o fenómenos que se caracterizan por estar sujetas a cambios pueden ser observadas, medidas, analizadas y controladas en el proceso de investigación.

### 2.3.1. Variable Independiente

Subida de los precios de los combustibles.

### 2.3.2. Variable Dependiente

Incidencia en el consumo de los habitantes del cantón El Triunfo.

Productos de la canasta básica que más han elevado su costo por subida del precio de los combustibles.

| Variable | Definición | Tipo De Medición e Indicador | Técnicas de Tratamiento de la Información | Resultados Esperados |
|---|---|---|---|---|
| Subida de los precios de los combustibles. | Combustible es cualquier material capaz de liberar energía cuando se oxida de forma violenta con desprendimiento de calor poco a poco, esto se usa para el funcionamiento de maquinarias. | Medición Cuantitativa<br><br>Indicador<br><br>Inflación | Información primaria, encuestas realizadas a los diferentes agentes económicos del sector comercial. | Examinar la subida del precio de los combustibles y como afecta al consumo. |
| Productos de la canasta básica que más han elevado su coso por subida del precio del combustible. | La canasta básica es una estimación de un grupo de alimentos principales, que a veces incluye otros artículos no alimentarios esenciales, y se ajusta según criterios tales como el porcentaje de gasto en alimentos para cada grupo de hogares (Incap.int, 2002). | Medición Cuantitativa<br><br>Indicador<br><br>Inflación<br><br>Cantidad de bienes y servicios IPC | Información primaria, Encuestas realizadas a los diferentes agentes económicos del sector comercial. | Examinar los productos que incrementaron su valor por la subida de los precios a los combustibles en el cantón. |
| Incidencia en el consumo de los habitantes del cantón el triunfo. | Consumo alimentario es la cantidad de comida o productos consumidos por un individuo. | Medición<br><br>Cualitativa<br><br>Indicador<br><br>Inflación<br><br>IPC | Información primaria, Encuestas realizadas a los diferentes agentes económicos del sector comercial. | Comparar el costo de la canasta básica con el salario mínimo obtenido por los habitantes. |

### 2.4 Población y Muestra

#### 2.4.1 Población

Una población se define como un grupo de individuos con una cantidad limitada o gran cantidad de datos para un sector, industria, empresa u organización que tiene como objetivo el proporcionar datos de investigación que contribuyan mediante la aplicación de una técnica. (Otzen & Manterola, 2017), afirman que: "Una población representa un grupo particular de personas para un posterior análisis de seguimiento y estudio de características internas y externas con la finalidad de proporcionar datos concretos al investigador".

Por ello para la presente investigación la población objeto de estudio serán las familias que vivan en el cantón El Triunfo, según el portal (Obras Públicas, 2017) se conoce la población que tiene el cantón El Triunfo que es de 59.636 mil habitantes al año 2020, de lo cual nos guiaremos por el dato de viviendas que tiene el cantón que es de 13.807 lo cual representara las familias.

#### 2.4.2 Muestra

Una muestra es un subconjunto de la población que tiene dos tipos de muestras: la probabilística y la no probabilista; La muestra probabilística es aquella en la que todos los elementos de la población cuentan con la misma probabilidad de selección mientras los criterios de selección no probabilísticos están definidos por el investigador. La muestra es una parte representativa de la población (López, 2018).

La técnica de muestreo que se usará para el presente artículo, es el de diseño probabilístico, con un muestreo aleatorio simple.

Para esto se utiliza la siguiente formula:

$$n = \frac{N * Z^2 * p * q}{d^2 * (N - 1) + Z^2 * p * q}$$

Donde:

➢ N= Total de la población.

➢ Z= 1.96, el nivel de confianza es del 95%.

➢ p= probabilidad de éxito, está determinado al 50%.

➢ q= 1-p probabilidad de fracaso, se considera el 50%.

➢ d= precisión de error, se usó el 10%.

$$n = \frac{13.807 * (1,96)^2 * 0,5 * (0,5)}{(0,10)^2 * (13807-1) + (1.96)^2 * (0,5) * (0,5)} = 95,38$$

El número de muestras consideradas en la presente investigación para realizar las encuestas dentro del cantón El Triunfo es de 95, las cuales serán tomadas de manera aleatoria.

## 2.5 Técnicas de Recolección de Datos

Los datos a recolectar que nos permitan obtener información de diferentes fuentes, con el fin de cumplir con los objetivos planteados para la redacción de este artículo se obtuvieron mediante:

### 2.5.1 Investigación Cuantitativa

- **Encuesta:** Es un proceso de la investigación descriptiva que consiste en recolectar información y datos mediante un cuestionario previamente elaborado, sin modificar el entorno en donde se realiza.

**Recursos bibliográficos**

- Sitios Web: artículos científicos, libros y tesis de grado.

| RECOLECCIÓN DE LA INFORMACIÓN | | | |
|---|---|---|---|
| *FUENTES* | | *TÉCNICAS* | |
| **PRIMARIAS** | Personas | **ENCUESTA** | Cuestionario |
| **SECUNDARIAS** | Material Digital | **OBSERVACIÓN** | A través de tecnologías de información y comunicación |

### 2.6 Estadística Descriptiva e Inferencial

Las técnicas que se van a emplear para determinar el incremento del precio de los combustibles y su incidencia en los productos de la canasta básica del cantón El Triunfo, se detallan a continuación:

### *2.6.1. Distribución de Frecuencias y Representaciones Gráficas*

- Gráficas de barras o pie (pastel): es una forma de representar un conjunto de datos o valores gráficamente con longitud proporcional al conjunto de datos que se presenta.

## CAPITULO III

### Resultados

**Descripción de la canasta básica y los efectos sociales que ocasiona el aumento de los precios.**

*Tabla 1. Grupos y subgrupos de la canasta familiar básica*

| Grupos y subgrupos de consumo | Costo actual en dólares |
|---|---|
| **Total** | **700,10** |
| **Alimentos y bebidas** | **236,60** |
| Cereales y derivados | 58,39 |
| Carne y preparaciones | 34,75 |
| Pescados y mariscos | 13,31 |
| Grasas y aceites comestibles | 9,20 |
| Leche, productos lácteos y huevos | 32,76 |
| Verduras frescas | 14,16 |
| Tubérculos y derivados | 14,14 |
| Leguminosas y derivados | 5,44 |
| Frutas frescas | 14,51 |
| Azúcar, sal y condimentos | 12,27 |
| Café, té y bebidas gaseosas | 6,93 |
| Otros productos alimenticios | 1,03 |
| Alimentos y bebidas consumidas fuera del hogar | 19,72 |
| **Vivienda** | **180,75** |
| Alquiler | 151,11 |
| Alumbrado y combustible | 16,59 |
| Lavado y mantenimiento | 11,60 |
| Otros artefactos del hogar | 1,46 |
| **Indumentaria** | **49,55** |
| Telas, hechuras y accesorios | 4,20 |
| Ropa confeccionada hombre | 23,22 |
| Ropa confeccionada mujer | 19,93 |
| Servicio de limpieza | 2,21 |
| **Misceláneos** | **233,19** |
| Cuidado de la salud | 104,16 |
| Cuidado y artículos personales | 16,63 |
| Recreo material de la lectura | 29,14 |
| Tabaco | 33,59 |
| Educación | 14,80 |
| Transporte | 34,88 |

*Fuente: (INEC, 2020)*

Este objetivo fue contestado a través de las preguntas 4,9 y 10 de la encuesta realizada a los habitantes dentro de la localización.

**4. ¿Debido al incremento de algunos productos el salario Básico le alcanza para cubrir sus necesidades de alimentación en su hogar?**

*Tabla 2. ¿Debido al incremento de algunos productos el salario Básico le alcanza para cubrir sus necesidades de alimentación en su hogar?*

| Respuestas | Frecuencia | Porcentaje |
|---|---|---|
| **Siempre** | 2 | 2% |
| **Casi siempre** | 26 | 27% |
| **Algunas veces** | 39 | 41% |
| **Muy pocas veces** | 20 | 21% |
| **Nunca** | 8 | 8% |
| **TOTAL** | **95** | 100% |

*Elaborado: Por el Autor, (2022).*

*Figura 1. ¿Debido al incremento de algunos productos el salario Básico le alcanza para cubrir sus necesidades de alimentación en su hogar?*

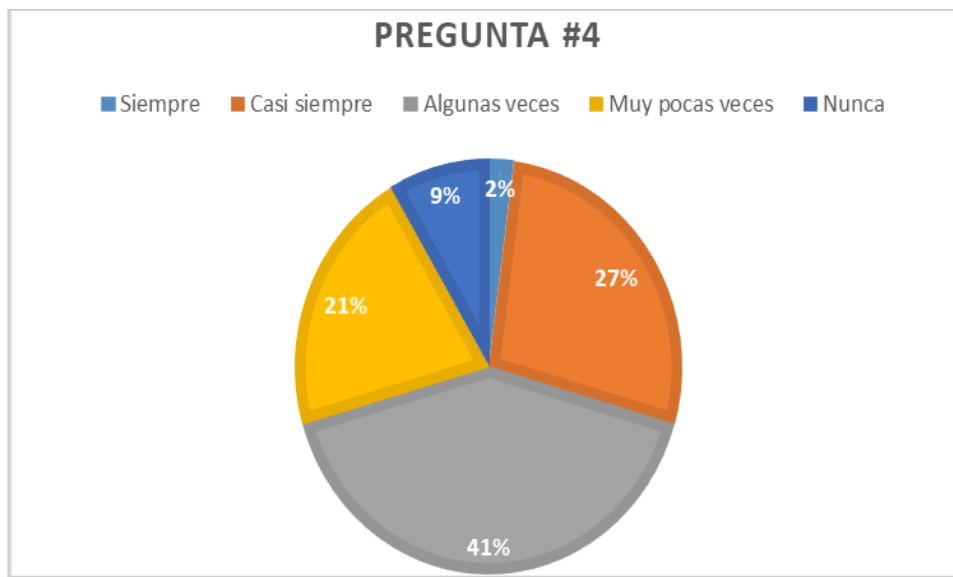

*Elaborado por: el Autor, 2022.*

**Análisis:**

Con base en el gráfico de las encuestas realizadas, podemos concluir que al 2% de la población encuestada, siempre le alcanza para cubrir sus necesidades de alimentación, un 27% casi siempre puede cubrir esta necesidad, un 41% de los encuestados respondió que algunas veces, un 21% y 8% considera que muy pocas veces y nunca pueden cubrir esta necesidad respectivamente.

**9. ¿Está usted de acuerdo con el valor que paga por los productos de la canasta básica?**

*Tabla 3. ¿Está usted de acuerdo con el valor que paga por los productos de la canasta básica?*

| Respuestas | Frecuencia | Porcentaje |
|---|---|---|
| Siempre | 5 | 5% |
| Casi siempre | 5 | 5% |
| Algunas veces | 20 | 21% |
| Muy pocas veces | 23 | 24% |
| **Nunca** | 42 | 44% |
| **TOTAL** | 95 | 100% |

*Elaborado: Por el Autor, (2022).*

*Figura 2. ¿Está usted de acuerdo con el valor que paga por los productos de la canasta básica?*

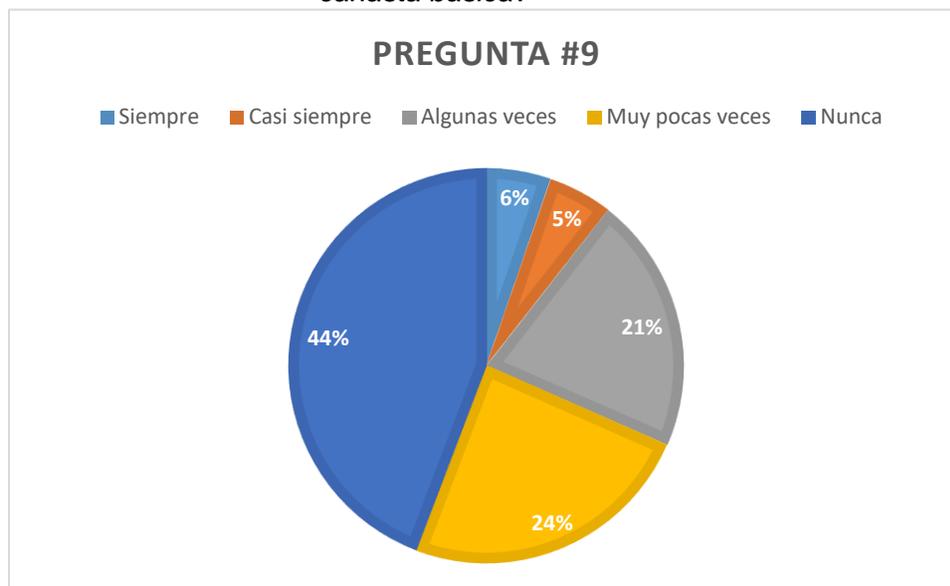

*Elaborado: Por el Autor, (2022).*

**Análisis:**

Con base a las encuestas realizadas podemos concluir que el 5% de encuestados siempre esta de acuerdo con el valor que paga por los productos de la CFB, otro 5% casi siempre, un 21% y un 24% algunas veces y muy pocas veces está de acuerdo respectivamente, mientras que un 44% nunca esta de acuerdo con el precio que pagan.

**10. ¿Considera usted que un aumento en el salario mejoraría su poder adquisitivo?**

*Tabla 4. ¿Considera usted que un aumento en el salario mejoraría su poder adquisitivo?*

| Respuestas | Frecuencia | Porcentaje |
|---|---|---|
| **Siempre** | 50 | 53% |
| **Casi siempre** | 24 | 25% |
| **Algunas veces** | 13 | 14% |
| **Muy pocas veces** | 6 | 6% |
| **Nunca** | 2 | 2% |
| **TOTAL** | 95 | 100% |

*Elaborado: Por el Autor, (2022).*

*Figura 3. ¿Considera usted que un aumento en el salario mejoraría su poder adquisitivo?*

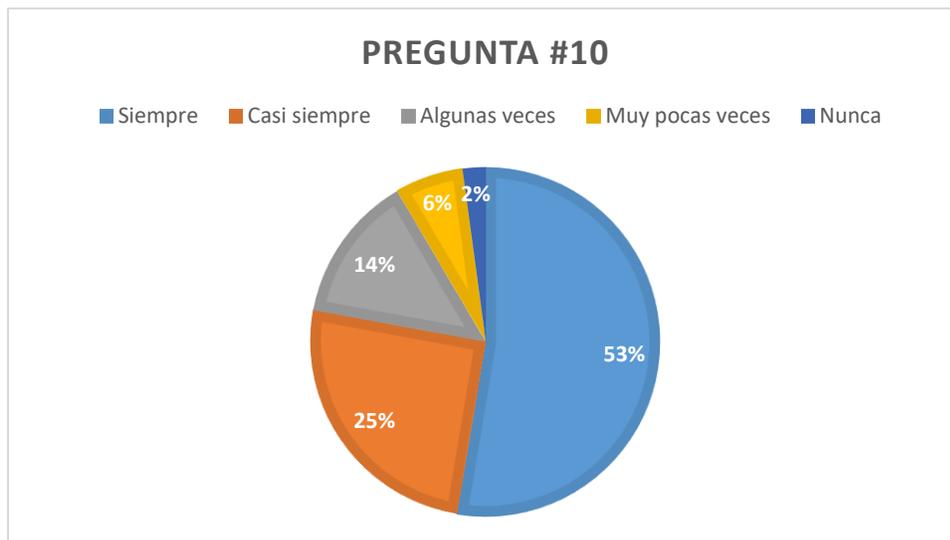

*Elaborado: Por el Autor, (2022).*

**Análisis:**

El 53% de los encuestados mencionó que siempre mejoraría su poder adquisitivo si se da un aumento en el SBU, el 25% considera que casi siempre mejoraría su poder adquisitivo al momento de darse un aumento en el SBU, el 14% considera que esto solo mejoraría algunas veces encontrándose de esta manera en un punto intermedio, por otro lado, existe un 6% que menciona que esto se daría muy pocas veces y por último esta que el 2% de la población encuestada menciona que nunca pasaría esto.

**Relación entre el precio del combustible Diésel y el precio de la canasta básica, en el periodo seleccionado.**

**Análisis de correlación**

El análisis de correlación consiste en un procedimiento estadístico para determinar si dos variables están relacionadas o no. El resultado del análisis es un coeficiente de correlación que puede tomar valores entre -1 y +1. (Alquicira, 2016) **A continuación, se presenta la tabla con el precio del combustible Diésel y el precio de la CFB, en el periodo seleccionado:**

*Tabla5. Precio del combustible Diésel y el precio de la CFB 2018-2022.*

| Año | X: Diésel/Galón/Precio | Y: Canasta básica/Precio |
|------|------------------------|--------------------------|
| **2018** | $2,61 | $715,16 |
| **2019** | $2,36 | $715,08 |
| **2020** | $1,18 | $710,08 |
| **2021** | $1,90 | $712,11 |
| **2022** | $1,75 | $751,04 |

Tomado y adaptado del (INEC, 2022); (PETROECUADOR, 2022), El autor, (2022).

**Gráfico estadístico de correlación entre el precio del Diesel y el precio de la canasta básica, en el periodo seleccionado:**

=COEF.DE.CORREL(C4:C8;D4:D8)

= -0,08990078

*Figura4. Correlación entre el precio del combustible y el precio de la canasta básica.*

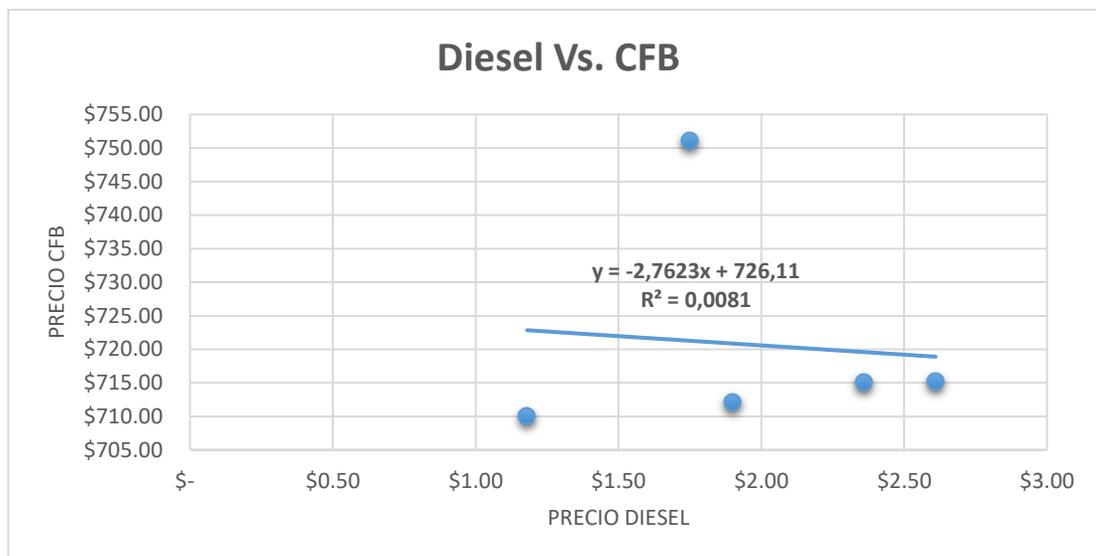

*Elaborado: Por el Autor, (2022).*

En la figura 4 se determinó, mediante una regresión lineal simple, que no existe correlación entre el precio del combustible Diesel, tomado como variable independiente, y el precio de la CFB establecida como variable dependiente, ya que el análisis arrojó un coeficiente de Pearson de -0,08990078, lo cual podemos determinar como una correlación negativa muy baja, es decir, el nivel de correlación entre las variables no es significativo.

Además, este objetivo fue contestado a través de las preguntas 3 y 5 de la encuesta realizada a los habitantes dentro de la localización.

1. **¿La subida de los precios de los combustibles afecta su poder adquisitivo?**

*Tabla 6. ¿La subida de los precios de los combustibles afecta su poder adquisitivo?*

| Respuestas | Frecuencia | Porcentaje |
|------------|-----------|-----------|
| Siempre | 61 | 64% |
| Casi siempre | 20 | 21% |
| Algunas veces | 8 | 8% |
| Muy pocas veces | 4 | 4% |
| Nunca | 2 | 2% |
| TOTAL | 95 | 100% |

*Elaborado: Por el Autor, (2022).*

*Figura 5. ¿La subida de los precios de los combustibles afecta su poder adquisitivo?*

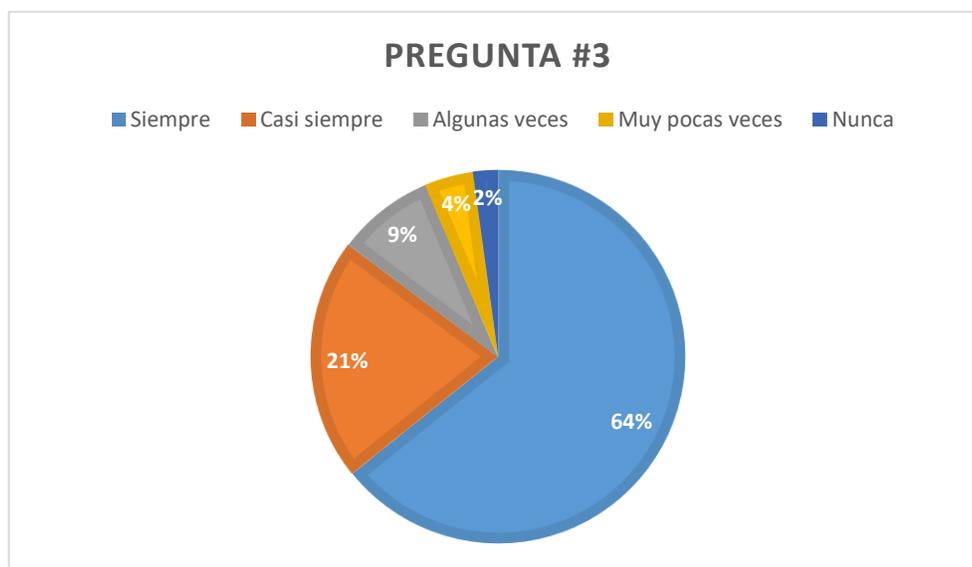

*Elaborado: Por el Autor, (2022).*

**Análisis:**

Basándonos en los resultados se puede concluir que el 64% de los encuestados considera que la subida del precio de los combustibles afectó su poder adquisitivo, un 21% casi siempre se vio afectado, el 8% contestó que algunas veces, el 4% y 2% muy pocas veces y nunca se vio afectado por este factor.

**5. ¿Considera usted que el aumento del precio en los combustibles ha influido en el alza de los precios de los productos en la canasta básica?**

*Tabla 7. ¿Considera usted que el aumento del precio en los combustibles ha influido en el alza de los precios de los productos en la canasta Básica?*

| Respuestas | Frecuencia | Porcentaje |
|---|---|---|
| **Siempre** | 64 | 67% |
| **Casi siempre** | 21 | 22% |
| **Algunas veces** | 6 | 6% |
| **Muy pocas veces** | 4 | 4% |
| **Nunca** | 0 | 0% |
| **TOTAL** | 95 | 100% |

*Elaborado: Por el Autor, (2022).*

*Figura 6. ¿Considera usted que el aumento del precio en los combustibles ha influido en el alza de los precios de los productos en la canasta Básica?*

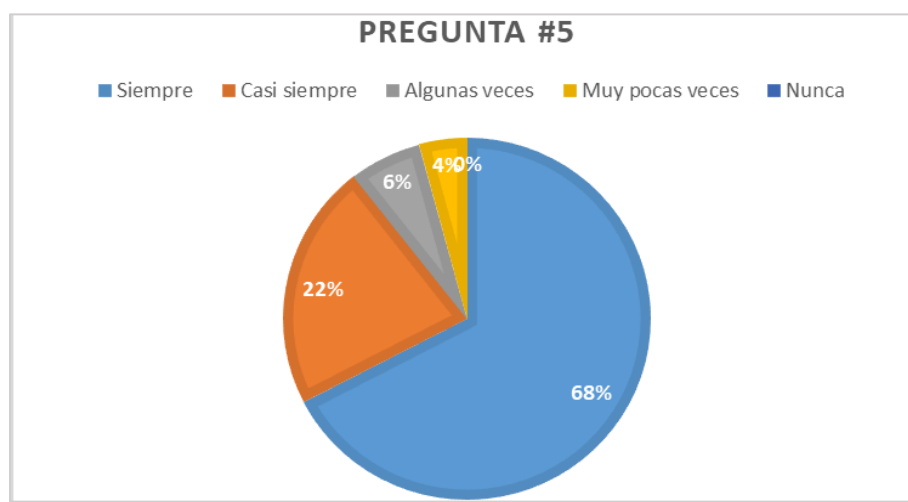

*Elaborado: Por el Autor, (2022).*

**Análisis:**

De acuerdo con los resultados de las encuestas realizadas, se puede concluir que el 68% de la muestra encuestada considera que el aumento en el combustible influye en los productos de la CFB, un 22% contestó que casi siempre influye, un 6% y 4% considera que algunas y muy pocas veces esto se da, y ningún encuestado considera que nunca.

**8. ¿Recibe alguna ayuda económica para cubrir sus gastos básicos en el hogar?**

*Tabla 8. ¿Recibe alguna ayuda económica para cubrir sus gastos básicos en el hogar?*

| Respuestas | Frecuencia | Porcentaje |
|---|---|---|
| Siempre | 1 | 1% |
| Casi siempre | 1 | 1% |
| Algunas veces | 5 | 5% |
| Muy pocas veces | 6 | 6% |
| Nunca | 82 | 86% |
| TOTAL | 95 | 100% |

*Elaborado: Por el Autor, (2022).*

*Figura 7. ¿Recibe alguna ayuda económica para cubrir sus gastos básicos en el hogar?*

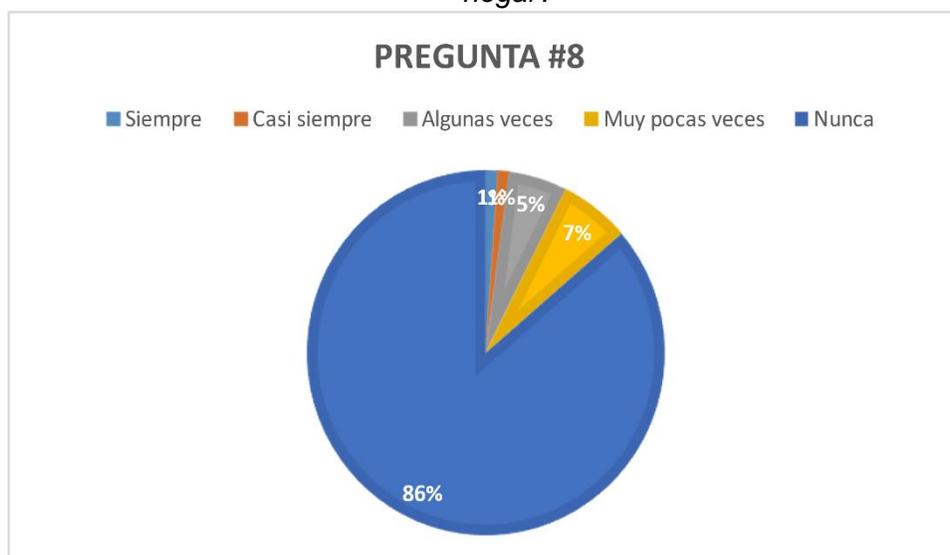

*Elaborado: Por el Autor, (2022).*

**Análisis:**

Para esta pregunta un 1% de los encuestados respondió que siempre y casi siempre recibe una ayuda para sus gastos básicos en el hogar, un 5% dijo algunas veces recibirla, un 6% y un 86% contestó que muy pocas veces y nunca reciben ayuda económica para gastos del hogar.

**Afectación directa que tuvo el alza de combustibles en el consumo básico de las familias del Cantón El Triunfo-Guayas.**

Este objetivo fue respondido a través de las preguntas 1, 2, 6 y 7 de la encuesta realizada a los habitantes dentro de la localización.

2. **Al realizar sus compras, ¿ha evidenciado usted un incremento de precios en los productos de la canasta básica?**

*Tabla 9. Al realizar sus compras, ¿ha evidenciado usted un incremento de precios en los productos de la canasta básica?*

| Respuestas | Frecuencia | Porcentaje |
|---|---|---|
| Siempre | 58 | 61% |
| Casi siempre | 27 | 28% |
| Algunas veces | 6 | 6% |
| Muy pocas veces | 0 | 0% |
| Nunca | 4 | 4% |
| TOTAL | 95 | 100% |

*Elaborado: Por el Autor, (2022).*

*Figura 8. Al realizar sus compras, ¿ha evidenciado usted un incremento de precios en los productos de la canasta básica?*

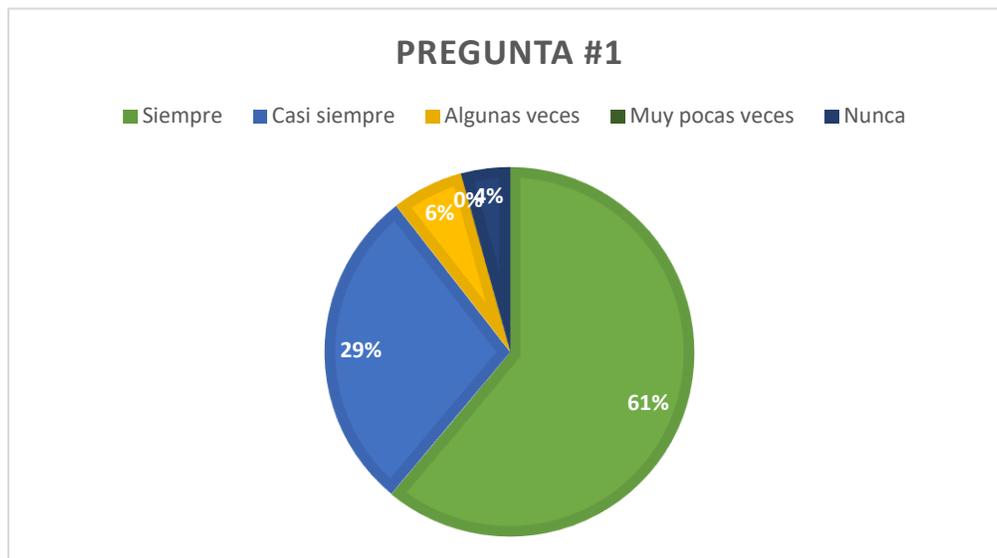

*Elaborado: Por el Autor, (2022).*

**Análisis:**

De acuerdo con los resultados obtenidos se puede concluir que el 61% siempre y 28% casi siempre han evidenciado un incremento en los productos de la CFB, un 6% algunas veces ha notado algún aumento y nunca ha evidenciado un aumento de los precios un 4% de la muestra encuestada.

3. **¿Ha tenido que reducir los productos de la canasta básica que usted compraba, debido al aumento de precios?**

*Tabla 10. ¿Ha tenido que reducir los productos de la canasta básica que usted compraba, debido al aumento de precios?*

| Respuestas | Frecuencia | Porcentaje |
|---|---|---|
| Siempre | 41 | 43% |
| Casi siempre | 16 | 17% |
| Algunas veces | 28 | 29% |
| Muy pocas veces | 9 | 9% |
| Nunca | 1 | 1% |
| TOTAL | 95 | 100% |

*Elaborado: Por el Autor, (2022).*

*Figura 9. ¿Ha tenido que reducir los productos de la canasta básica que usted compraba, debido al aumento de precios?*

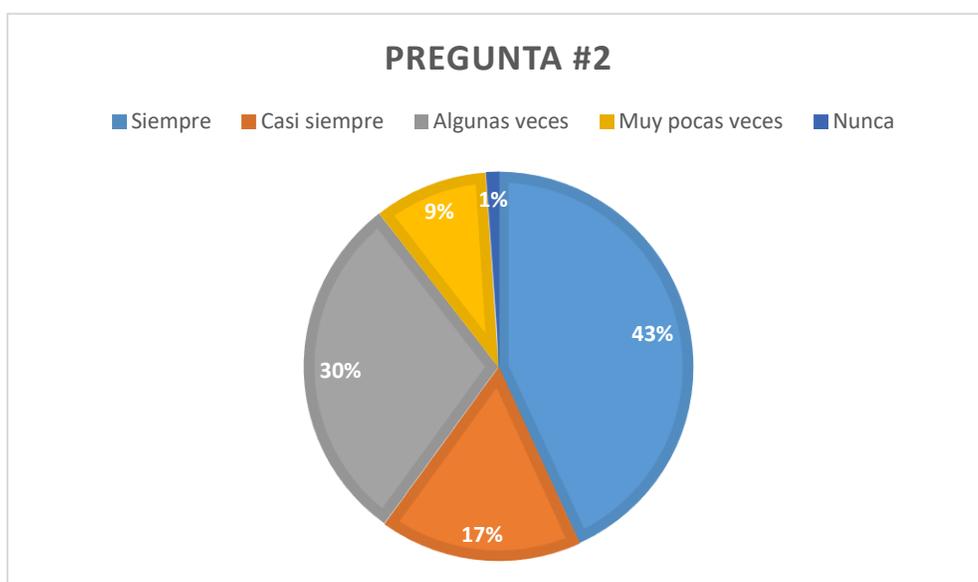

*Elaborado: Por el Autor, (2022).*

**Análisis:**

Con base en los resultados se observa que el 43% y 17% siempre y casi siempre han tenido que reducir los productos de la canasta básica que compraban, debido al aumento de precios, el 29% algunas veces tuvo que reducir sus compras, mientras que un 9% y 1% muy pocas veces y nunca se han visto en la necesidad de hacerlo, según lo contestado.

6. **¿Considera usted que el incremento del precio del combustible afecta su presupuesto familiar?**

*Tabla 11. ¿considera usted que el incremento del precio del combustible afecta su presupuesto familiar?*

| Respuestas | Frecuencia | Porcentaje |
|:---:|:---:|:---:|
| Siempre | 55 | 58% |
| Casi siempre | 24 | 25% |
| Algunas veces | 8 | 8% |
| Muy pocas veces | 6 | 6% |
| Nunca | 2 | 2% |
| TOTAL | 95 | 100% |

*Elaborado: Por el Autor, (2022).*

*Figura 10. ¿Considera usted que el incremento del precio del combustible afecta su presupuesto familiar?*

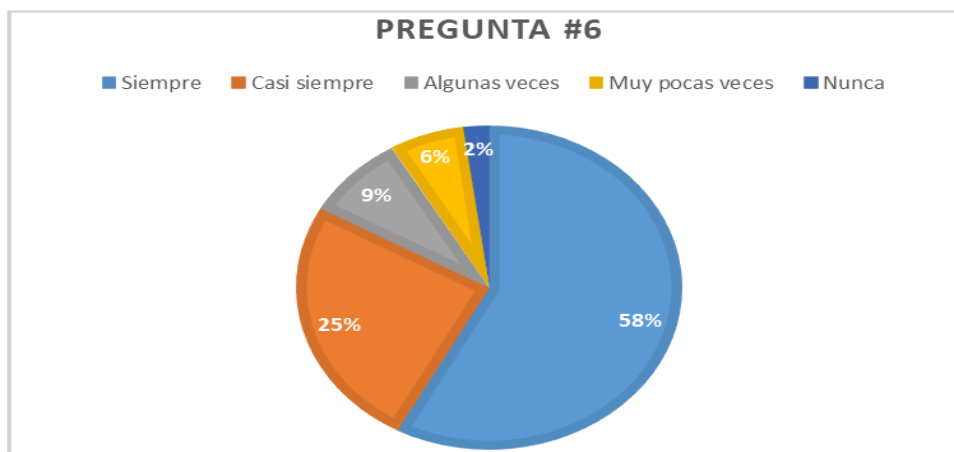

*Elaborado: Por el Autor, (2022).*

**Análisis:**

De acuerdo con nuestra gráfica, podemos concluir que al 58% de la población, siempre el alza del precio de los combustibles le ha afectado en su presupuesto mensual, un 25% considera que casi siempre se vio afectado, 8% y 6% respondieron que algunas veces y muy pocas veces respectivamente, mientras que al 2% nunca le afectó.

**7. ¿Usted ha dejado de consumir ciertos productos ya que subieron de costo en la actualidad?**

*Tabla 12. ¿Usted ha dejado de consumir ciertos productos ya que subieron de costo en la actualidad?*

| Respuestas | Frecuencia | Porcentaje |
|:---:|:---:|:---:|
| **Siempre** | 37 | 39% |
| **Casi siempre** | 21 | 22% |
| **Algunas veces** | 21 | 22% |
| **Muy pocas veces** | 12 | 13% |
| **Nunca** | 4 | 4% |
| **TOTAL** | 95 | 100% |

*Elaborado: Por el Autor, (2022).*

*Figura 11. ¿Usted ha dejado de consumir ciertos productos ya que subieron de costo en la actualidad?*

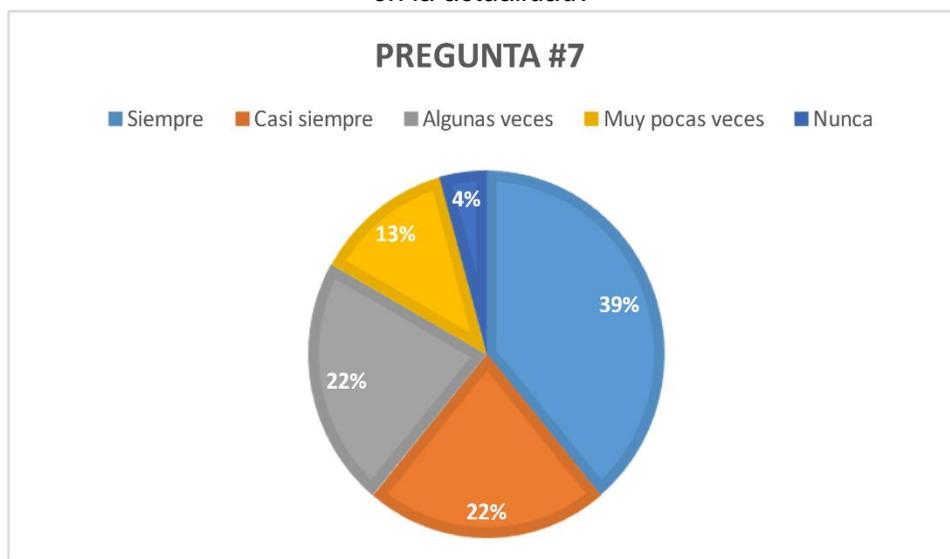

*Elaborado: Por el Autor, (2022).*

**Análisis:**

En el cantón El Triunfo, el 39% de las personas encuestadas contestaron que han dejado de consumir ciertos productos ya que subieron de costo en la actualidad, el 22% casi siempre y algunas veces lo hicieron, muy pocas veces dejaron de consumir productos debido a que aumentaron de precio un 13%, mientras que el 4% nunca dejaron de hacerlo.

**CONCLUSIONES Y RECOMENDACIONES**

**Conclusiones**

- A lo largo de la realización del presente proyecto, se obtuvo que la canasta básica está compuesta por secciones como alimentos y bebidas que engloban productos como verduras, legumbres, carnes, entre otros; también dentro de la sección se encuentra la vivienda que incluye alquiler, alumbrado, combustible entre otras; la indumentaria que es ropa y accesorios y por último los misceláneos que incluye salud, artículos personales, educación. Los efectos sociales que podemos determinar, como consecuencia del incremento de los combustibles, son un aumento del costo de vida, debido a que la mayoría de ecuatorianos no cuenta con los ingresos suficientes para satisfacer la canasta básica que se encuentra valorada en USD $751,04 y el salario básico de los ecuatorianos está en USD $ 425.00 por ende el mismo no es suficiente para poder cubrir las necesidades establecidas en la canasta básica.

- En base a la correlacion realizada se llego a la conclusion de que no hay una relacion del incremento del precio de los combustibles con la canasta basica familiar, ya que mediante la resolucion del calculo del coeficioente de Pearson nos dio un valor de -0,08990078, lo cual podemos determinar como una correlación negativa muy baja.

- El alza de combustibles en el consumo básico de las familias del Cantón El Triunfo, se evidenció de manera directa según las encuestas realizadas a la población, ya que un gran porcentaje de encuestados expresó que sí cambiaron sus hábitos de consumo, a causa del incremento en sus precios, dado que se vio afectado su poder adquisitivo con referencia a los productos de la canasta básica, un gran porcentaje de familias redujo su consumo o eliminó productos de primera necesidad de su lista de compras.

**Recomendaciones**

Para futuros estudios que pudieran complementar el presente trabajo se establecen las siguientes recomendaciones:

- Determinar si otras variables influyeron en el alza de precios de los productos de la canasta básica evidenciada a lo largo del presente año.

- Levantar información en los comercios, tiendas y supermercados de "El Triunfo", acerca de cada producto que integra la canasta básica, de manera que permita determinar un precio exacto de esta en la ciudad, para poder ampliar las posibilidades de visualizar el nivel de impacto del factor combustible en la canasta básica por Cantón.

- Debido a que se evidenció que el incremento de precio de algunos productos de la canasta básica, se dio por la subida del precio de los combustibles, se recomienda contar con reformas o leyes por parte de las autoridades gubernamentales, o del GAD municipal, que puedan regular con mayor efectividad los precios de los productos de la canasta básica, es decir, vigilar que no haya especulación de precios.

# BIBLIOGRAFÍA

(2008). En *CONTITUCIÓN DE LA REPÚBLICA DEL ECUADOR.* Obtenido de http://www.acnur.org/fileadmin/Documentos/BDL/2008/6716.pdf

Aguilar, I. (07 de 2019). *Repositorio de la Universidad Pontificia Comillas.* Obtenido de Universidad Pontificia Comillas: https://repositorio.comillas.edu/xmlui/handle/11531/31826

Bowen, A., & Chimbolema, B. (2021). *Incremento del precio del combustible diésel y su incidencia en el precio de la canasta básica en Guayaquil período 2015-2020.* Universidad de Guayaquil, Guayaquil. Obtenido de http://repositorio.ug.edu.ec/bitstream/redug/55777/1/BOWEN%20RODRIGUEZ%20ARIANA%20%26%20CHIMBOLEMA%20PILAMUNGA%20BEATR%c3%8dZ.pdf

Brito , L., Quito, M., Rodríguez , E., & Uriguen, P. (2021). Evolución del precio de la canasta básica del Ecuador. Análisis del periodo 2000 – 2019. *Revista Científica y Tecnológica UPSE, 08*(02), 9. Recuperado el 31 de 05 de 2022, de Repositorio Revista Científica y Tecnológica UPSE: https://incyt.upse.edu.ec/ciencia/revistas/index.php/rctu/article/view/551

Bure, M. C., Guerrero, E. L., Aguirre, P. A., & Gaona, L. F. (28 de Diciembre de 2021). Evolución del precio de la canasta básica del Ecuador. Análisis del periodo 2000 –2019. *Revista Científica y Tecnológica UPSE, Vol 8*, 59-67. Obtenido de https://incyt.upse.edu.ec/ciencia/revistas/index.php/rctu/article/view/551/520

Chicaiza, P. (2019). Las políticas de eliminación en los subsidios de los combustibles fósiles y su relación con la inflación del Ecuador. *Universidad técnica de Ambato .* Obtenido de http://repositorio.uta.edu.ec/bitstream/123456789/30614/1/T4694e.pdf

Coba, G. (6 de Mayo de 2022). La inflación llegó a 2,89% en abril de 2022, según el INEC. *Primicias*, págs. https://www.gob.mx/agricultura/articulos/la-canasta-basica-que-es-y-para-que-sirve-189256.

Davila, D. (2022). Conflicto Rusia-Ucrania podría encarecer algunos productos de la canasta básica en Ecuador, alerta economista. *Radio Pichincha*, 1. Obtenido de

https://www.pichinchacomunicaciones.com.ec/conflicto-rusia-ucrania-podria-encarecer-algunos-productos-de-la-canasta-basica-en-ecuador-alerta-economista/

Escribano, G. (2019). *Ecuador y los subsidios a los combustibles.* Real Instituto elcano. Obtenido de https://media.realinstitutoelcano.org/wp-content/uploads/2021/11/ari110-2019-escribano-ecuador-y-los-subsidios-a-los-combustibles.pdf

*esimpact.* (2021). Obtenido de https://www.esimpact.org/impacto-social/

España, S. (2 de Octubre de 2019). *Ecuador elimina los subsidios a la gasolina para corregir sus estrecheces fiscales*. Obtenido de El País: https://elpais.com/internacional/2019/10/02/america/1570042474_164745.html

España, S. (16 de octubre de 2021). El costo de la vida se encarece en Ecuador ante la subida del petróleo y la reactivación económica. Obtenido de El País: https://elpais.com/america/economia/2021-10-17/el-costo-de-la-vida-se-encarece-en-ecuador-ante-la-subida-del-petroleo-y-la-reactivacion-economica.html

Freire, C., Mayorga, F., Vayas, T., & Sánchez, A. M. (12 de 2020). *Universidad Técnica de Ambato*, Digital. Recuperado el 31 de 05 de 2022, de Blogs Cedia: https://blogs.cedia.org.ec/obest/wp-content/uploads/sites/7/2020/12/Canasta-Familiar-Basica-Ecuatoriana.pdf

Gronneberg, I. (2022). ¿Está el Ecuador preparado para los efectos de la guerra? *Vistazo*, 1. Obtenido de https://www.vistazo.com/opinion/columnistas/inty-gronneberg/esta-el-ecuador-preparado-para-los-efectos-de-la-guerra-AD1423525

Guzmán, C. (15 de Enero de 2020). *PQS*. Obtenido de https://pqs.pe/actualidad/economia/que-es-la-canasta-basica-para-que-sirve/

Icaria, E. (2005). *Escola de Cultura de Pau*. Obtenido de https://escolapau.uab.cat/conflictes-armats/#:~:text=Entendemos%20por%20conflicto%20armado%20%E2%80%9Ctodo,de%20destrucci%C3%B3n%2C%20provocan%20m%C3%A1s%20de

INEC. (2017). Obtenido de www.ecuadorencifras.gob.ec/institucional/home/.

INEC. (2020). *Instituto Nacional de Estadísticas y Censo*. (INEC, Editor, & INEC, Productor) Obtenido de Instituto Nacional de Estadísticas y Censo: https://www.ecuadorencifras.gob.ec/institucional/home/#

Menger, C. (s.f.). Economía y bienestar económico. En *Principios de economia política* (págs. 69-101). Barcelona: Ediciones Orbis. Obtenido de http://economia.unam.mx/profesores/blopez/bienestar-menger.pdf

Organización Internacional del Trabajo. (s.f.). Obtenido de https://www.ilo.org/global/topics/wages/minimum-wages/definition/lang--es/index.htm

Orozco, M. (05 de Mayo de 2022). *primicias.ec.* Obtenido de https://www.primicias.ec/noticias/economia/importacion-combustibles-aumento-subsidios-ecuador/

PETROECUADOR, E. (10 de Octubre de 2020). *EP PETROECUADOR.* Obtenido de http://www.eppetroecuador.ec/?p=9338

Ramírez, & Dionicia. (Marzo de 2020). *Repositorio Universidad tecnica de Ambato.* Obtenido de http://repositorio.uta.edu.ec/bitstream/123456789/34352/1/T5233e.pdf

Rivera, A. D., & Villacis, F. J. (2019). *Revista Observatorio de la Economía.* Obtenido de https://www.eumed.net/rev/oel/2019/07/inflacion-canasta-basica.html

Roca, R. (s.f.). *Teorías de la inflación* . (ACADEMIA, Ed.) Lima, Perú. Obtenido de https://d1wqtxts1xzle7.cloudfront.net/26250553/roca(1999)teoriasinflacion-with-cover-page-v2.pdf?Expires=1654054281&Signature=bd-VmFgQQB~Jx40Wh5sV4P4nIxUV4ACLNnkzWR4CzAQlZl5weDCn-sdXMer9xdmsI-z4O22wK9nuXQdbtyMqT~MCa3-DmZ9p6Sp9LPv8lkrgyxPNGatD6zW6BLGDhfkRo

Ruiz, J. A. (2021). *IMPACTO ECONÓMICO, POLÍTICO Y SOCIAL EN EL SUBSIDIO DEL COMBUSTIBLE Y GAS LICUADO DE PETRÓLEO (GLP) EN EL ECUADOR.*

Salazar, & Zurita. (2017). *https://www.dspace.espol.edu.* Obtenido de https://www.dspace.espol.edu.ec/bitstream/123456789/2107/1/4204.pdf

Schuldt, J., & Acosta, A. (2013). *Inflacion.* Obtenido de https://books.google.es/books?hl=es&lr=&id=PDO2ZXU28UYC&oi=fnd&pg=PA7&dq=inflaci%C3%B3n+en+el+ecuador&ots=Qby_a98Ei-&sig=zUrDTRdIVtgRfKgDMtbOR4_T7nQ#v=onepage&q=inflaci%C3%B3n%20en%20el%20ecuador&f=true

Suárez, M. (2021). Eliminar los subsidios a los combustibles y proteger a los más pobres sí es posible. *Gestion Digital.* Obtenido de https://www.revistagestion.ec/economia-y-finanzas-analisis/eliminar-los-subsidios-los-combustibles-y-proteger-los-mas-pobres-si

Suarez, M. (06 de 02 de 2022). *Revista Gestión*, Digital. (R. Gestión, Editor) Recuperado el 31 de 05 de 2022, de Revista Gestión: https://www.revistagestion.ec/analisis-economia-y-finanzas/la-canasta-familiar-esta-mas-que-cubierta-pero-para-pocos

*traders.studio.* (11 de mayo de 2022). Obtenido de https://traders.studio/definicion-de-costo-de-vida/

Westreicher, G. (8 de agosto de 2020). *Economipedia.* Obtenido de https://economipedia.com/definiciones/combustible.html

**ANEXOS**

**ANEXO1.** Ubicación del cantón El Triunfo

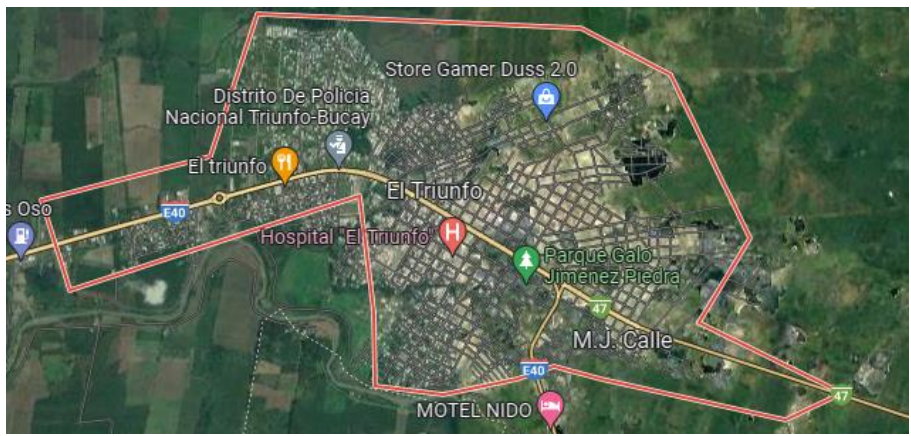

**Fuente:** Google Maps, 2022

**ANEXO2.** Evidencia fotográfica de la recolección de información mediante encuestas.

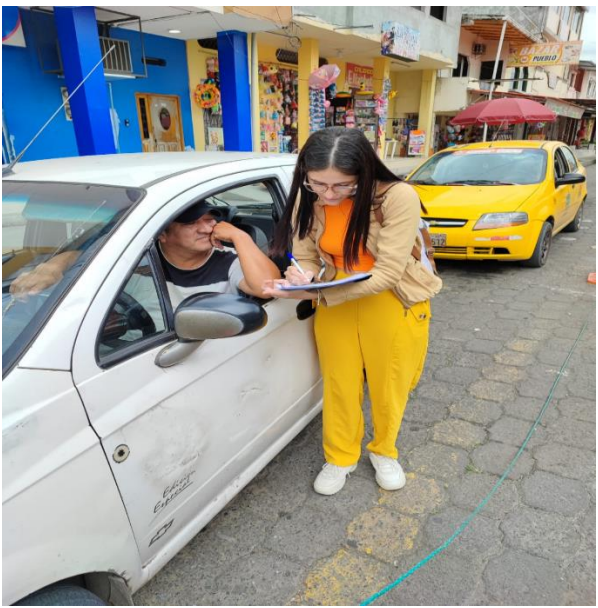
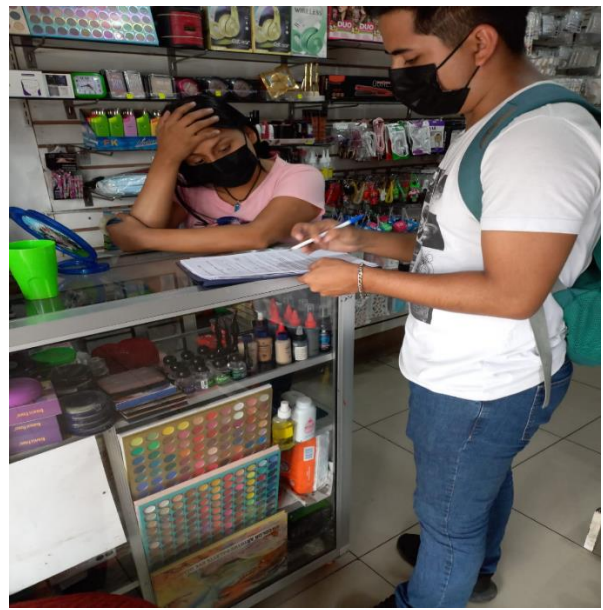

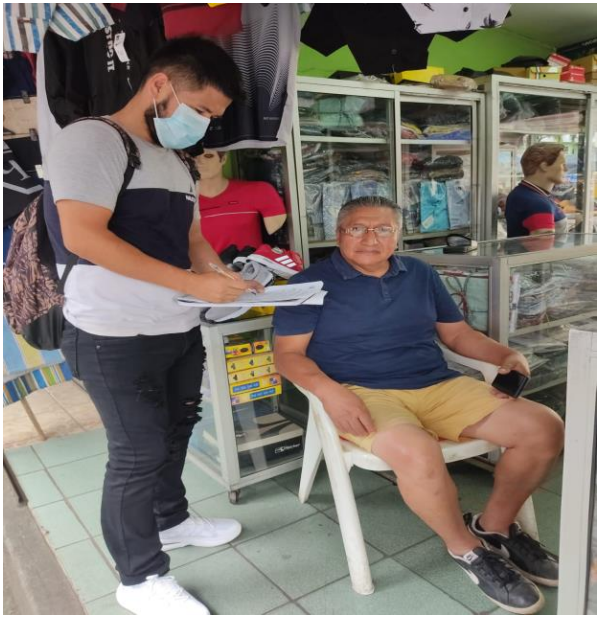

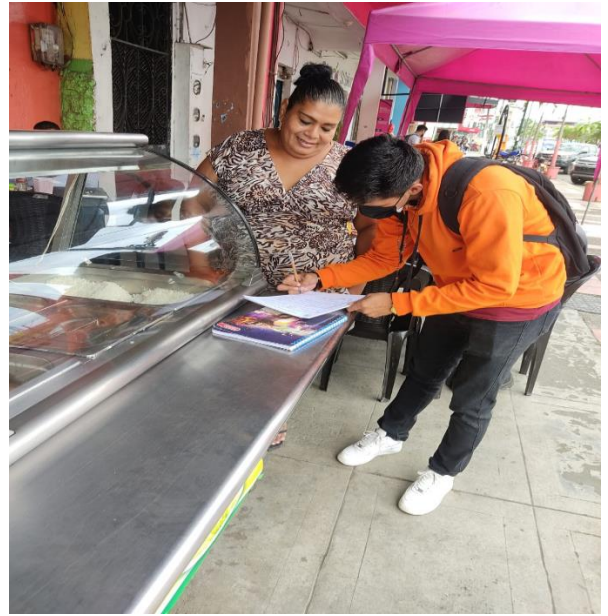

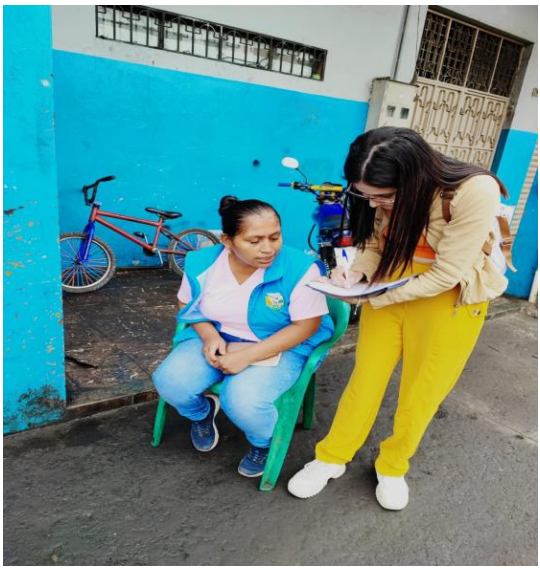

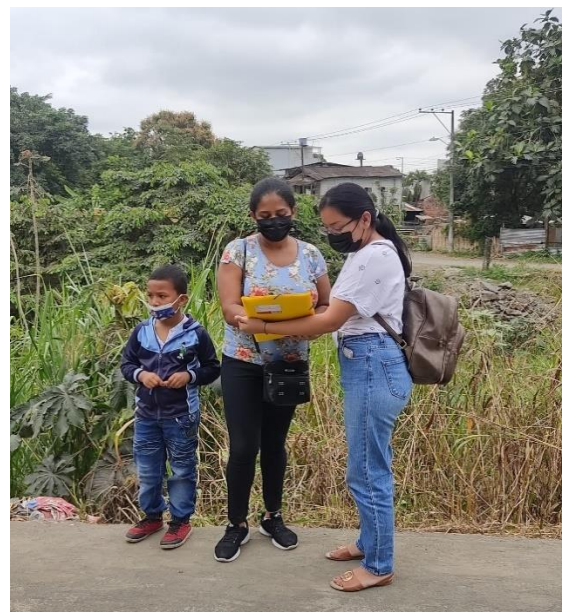

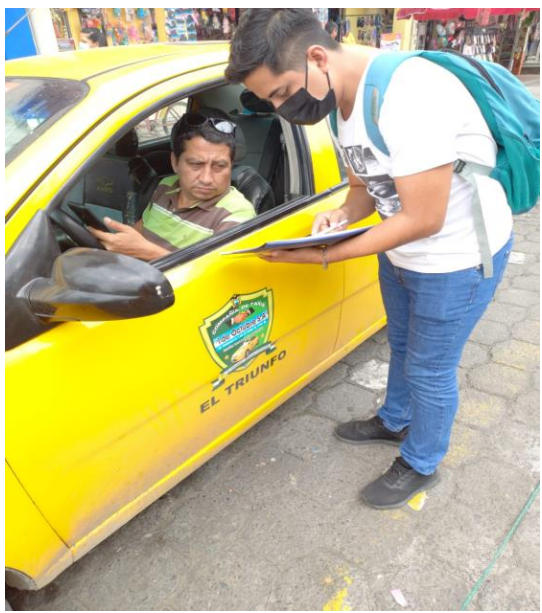 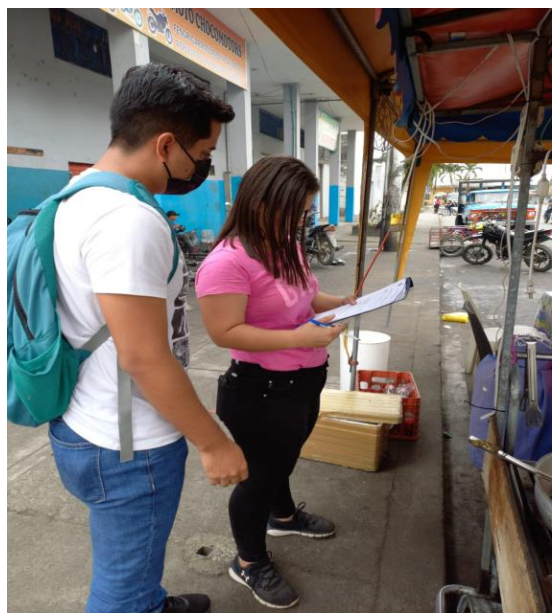

**Fuente:** Los Autores, 2022

**ANEXO3.** Encuesta

**Encuesta sobre "Incremento del precio de los combustibles y su impacto en la canasta básica"**

| Nombre: | | Edad: | | |
|---|---|---|---|---|
| Cedula: | | Sexo: | | |
| **Siempre** | Casi Siempre | Algunas veces | Muy pocas veces | Nunca |
| **5** | 4 | 3 | 2 | 1 |

**Coloque un visto dentro del parámetro que coincidan con su elección**

| N | Pregunta | Escala de Evaluación | | | | |
|---|---|---|---|---|---|---|
| | | 5 | 4 | 3 | 2 | 1 |
| **1** | Al realizar sus compras, ¿ha evidenciado usted un incremento de precios en los productos de la canasta básica? | ☐ | ☐ | ☐ | ☐ | ☐ |
| **2** | ¿Ha tenido que reducir los productos de la canasta básica que usted compraba, debido al aumento de precios? | ☐ | ☐ | ☐ | ☐ | ☐ |
| **3** | ¿La subida de los precios de los combustibles afecta su poder adquisitivo? | ☐ | ☐ | ☐ | ☐ | ☐ |
| **4** | ¿Debido al incremento de algunos productos el salario básico le alcanza para cubrir sus necesidades de alimentación en su hogar? | ☐ | ☐ | ☐ | ☐ | ☐ |
| **5** | ¿Considera usted que el aumento del precio en los combustibles ha influido en el alza de los precios de los productos en la canasta básica? | ☐ | ☐ | ☐ | ☐ | ☐ |
| **6** | ¿Considera usted que el incremento del precio del combustible afecta su presupuesto familiar? | ☐ | ☐ | ☐ | ☐ | ☐ |
| **7** | ¿Usted ha dejado de consumir ciertos productos ya que subieron de costo en la actualidad? | ☐ | ☐ | ☐ | ☐ | ☐ |
| **8** | ¿Recibe alguna ayuda económica para cubrir sus gastos básicos en el hogar? | ☐ | ☐ | ☐ | ☐ | ☐ |
| **9** | ¿Está usted de acuerdo con el valor que paga por los productos de la canasta básica? | ☐ | ☐ | ☐ | ☐ | ☐ |
| **10** | Considera usted que un aumento en el salario mejoraría su poder adquisitivo. | ☐ | ☐ | ☐ | ☐ | ☐ |